\documentclass[aps,prd,onecolumn,nofootinbib,groupedaddress,superscriptaddress,10pt]{revtex4}  
\usepackage{graphicx}
\usepackage{epstopdf}
\usepackage{amsmath}
\usepackage{amsfonts}
\usepackage{amssymb}
\usepackage{appendix}
\usepackage{comment}
\usepackage{bbold}
\usepackage{color}
\usepackage{slashed}
\usepackage{subfigure}
\usepackage{setspace}
\usepackage{footnote}
\usepackage{multirow}
\usepackage{longtable}
\usepackage{braket}
\usepackage[letterpaper, portrait, top=0.75in, left=0.5in, right=0.4in, bottom=0.4in]{geometry}
\usepackage[colorlinks=true,allcolors=purple]{hyperref}
\usepackage[capitalize]{cleveref}

\newcommand{\beq}{\begin{equation}}
\newcommand{\eeq}{\end{equation}}

\usepackage{hyperref}
\definecolor{nicered}{rgb}{0.7,0.1,0.1}
\definecolor{nicegreen}{RGB}{53,170,102}
\definecolor{niceblue}{RGB}{57,94,221}
\definecolor{nicepurple}{RGB}{127, 38, 222}
\hypersetup{colorlinks,
			citecolor      = nicepurple, 
			linkcolor       = niceblue, 
			urlcolor        = nicegreen, 
			anchorcolor = niceblue}

\begin{document}

\singlespacing

{\hfill MI-HET-846}

\title{On T-Invariance Violation in Neutrino Oscillations and Matter Effects}

\author{Olivia M. Bitter} 
\affiliation{Northwestern University, Department of Physics \& Astronomy, 2145 Sheridan Road, Evanston, IL 60208, USA}
\author{Andr\'{e} de Gouv\^{e}a} 
\affiliation{Northwestern University, Department of Physics \& Astronomy, 2145 Sheridan Road, Evanston, IL 60208, USA}
\author{Kevin J. Kelly}
\affiliation{Mitchell Institute for Fundamental Physics and Astronomy, Department of Physics and Astronomy, Texas A\&M University, College Station, TX 77843, USA}

\begin{abstract}
\vspace{4mm}

We investigate the impact of matter effects on T (time-reversal)-odd observables, making use of the quantum-mechanical formalism of neutrino-flavor evolution. We attempt to be comprehensive and pedagogical. Matter-induced T-invariance violation (TV) is qualitatively different from, and more subtle than, matter-induced CP (charge-parity)-invariance violation. If the matter distribution is symmetric relative to the neutrino production and detection points, matter effects will not introduce any new TV. However, if there is intrinsic TV, matter effects can modify the size of the T-odd observable. On the other hand, if the matter distribution is not symmetric, there is genuine matter-induced TV. For Earth-bound long-baseline oscillation experiments, these effects are small. This remains true for unrealistically-asymmetric matter potentials (for example, we investigate the effects of ``hollowing out'' 50\% of the DUNE neutrino trajectory). More broadly, we explore consequences, or lack thereof, of asymmetric matter potentials on oscillation probabilities. While fascinating in their own right, T-odd observables are currently of limited practical use, due in no small part to a dearth of intense, well-characterized, high-energy electron-neutrino beams. Further in the future, however, intense, high-energy muon storage rings might become available and allow for realistic studies of T invariance in neutrino oscillations. 

\end{abstract}

\maketitle

\section{Introduction}
\label{sec:Intro}
\setcounter{equation}{0}

The current and next generation of accelerator-based, long-baseline neutrino-oscillation experiments is well positioned to measure appearance and disappearance channels involving muon-type neutrinos $\nu_{\mu}$ and antineutrinos $\bar{\nu}_{\mu}$ in the initial state \cite{NOvA:2021nfi, T2K:2021xwb, Hyper-Kamiokande:2018ofw,DUNE:2020ypp}. The reason is these make use of, mostly, pion decays in flight to produce neutrino beams. Armed with these different oscillation channels, these so-called super-beam experiments can address the status of CP invariance in the neutrino sector by comparing the oscillations of $\nu_{\mu}$ to electron-type neutrinos $\nu_e$ to  those of $\bar{\nu}_{\mu}$ to electron-type antineutrinos $\bar{\nu}_{e}$: $P_{\mu e}$ versus $P_{\bar{\mu}\bar{e}}$.

The fact that the neutrinos in long-baseline oscillation experiments usually traverse a non-trivial amount of ordinary matter (with densities $\rho = \mathcal{O}$(few) g/cm$^3$) renders the study of CP invariance in neutrino oscillations more involved \cite{Wolfenstein:1977ue,Mikheyev:1985zog,Kuo:1989qe}. The reason is very easy to understand. In vaccum $P_{\mu e}$ and $P_{\bar{\mu}\bar{e}}$ are related by a CP transformation: schematically
\begin{align*}
{\rm CP}(P^{\rm vaccum}_{\mu e}) = P^{\rm vacuum}_{\bar{\mu}\bar{e}}.
\end{align*}
When the neutrinos propagate through matter,
\begin{align*}
{\rm CP}(P^{\rm matter}_{\mu e}) = P^{\rm \overline{matter}}_{\bar{\mu}\bar{e}},
\end{align*}
where $\rm \overline{matter}$ indicates the replacement of the ordinary matter traversed by the neutrino with the corresponding antimatter. Given our inability to build experiments in an anti-Earth, we are stuck with comparing $P^{\rm matter}_{\mu e}$ and $P^{\rm matter}_{\bar{\mu}\bar{e}}$. In summary, when neutrinos oscillate in matter, we cannot perform a direct, proper test of CP invariance, but must rely on an improper test of CP invariance and use our knowledge of the dynamics of neutrino flavor evolution in matter to study CP invariance in the lepton sector. 

The study of T invariance in neutrino oscillations is less prevalent in the neutrino oscillation literature for a good, practical reason. T invariance relates, for example, $P_{\mu e}$ with $P_{e\mu}$, and most intense, well characterized sources of electron-type neutrinos or antineutrinos (the Sun, nuclear reactors, etc.) yield neutrinos whose energies are low enough that the detection of muon-type neutrinos via charged-current weak interactions is kinematically forbidden. In the intermediate future, however, intense, high-energy muon storage rings -- `neutrino factories' -- might be available and would yield intense, high-energy beams of $\nu_{\mu}$ and $\bar{\nu}_e$ (or $\bar{\nu}_{\mu}$ and $\nu_e$, depending on the charge of the stored muons) and allow for realistic studies of T invariance in neutrino oscillations (for overviews and recent studies, see \cite{IDS-NF:2011swj,Delahaye:2018yfq,Bogacz:2022xsj,Denton:2024glz,Kitano:2024kdv,Ota:2001cz}). 

Similar to the case of CP-invariance studies, studies of T invariance in neutrino oscillations are also impacted by matter effects \cite{Kuo:1987km,Krastev:1988yu,Toshev:1989vz,Arafune:1996bt,Miura:2001pia,Akhmedov:2001kd,Parke:2000hu,Petcov:2018zka,Bernabeu:2019npc,Schwetz:2021cuj,Yasuda:2024lwg,Chatterjee:2024jzt,Lipari:2001ds,Schwetz:2021thj,Yokomakura:2000sv,deGouvea:2000un,Bernabeu:2018twl,Koike:1999tb,Sato:1999wt,Bilenky:1997dd,Koike:1999hf,Barger:1999hi,Yokomakura:2002av,Blom:2004bk,Rout:2017udo,Minakata:2002qi,Bernabeu:2018use,Xing:2013uxa,Henley:2011mu,Ohlsson:2003nb,Ota:2001cz,Akhmedov:1999uz,Koike:2000jf,Jacobson:2003wc,Bernabeu:1999ct,Bernabeu:1999gr,Cabibbo:1977nk,Yasuda:1999uv,Koike:1997dh,Singh:2016dpd,Minakata:2002qe,Harrison:1999df,Kuo:1989qe}. In vacuum, $P_{\mu e}$ and $P_{e\mu}$ are related by a T transformation: schematically, ${\rm T}(P^{\rm vaccum}_{\mu e}) = P^{\rm vacuum}_{e\mu}$. In some detail, the reason for this is as follows. The discrete T transformation relates the process where a neutrino is produced as a $\nu_\mu$ at position $a$ and later detected as a $\nu_e$ at position $b$ to the process where a $\nu_e$ is produced at position $b$ and detected later as a $\nu_\mu$ at position $a$. In vacuum, all points in spacetime are equivalent and the $a\to b$ trajectory is indistinguishable from the $b\to a$ trajectory and hence we write
\begin{align*}
{\rm T}(P^{\rm vaccum}_{\mu e}) = P^{\rm vacuum}_{e\mu},
\end{align*}
independent from the specific coordinates of the production and detection points (as long as the baselines are the same). In matter, one needs to be more careful and remember that a T transformation translates into 
\begin{align*}
{\rm T}(P^{\rm matter}_{\mu@a\to e@b}) = P^{\rm matter}_{e@b\to\mu@a}. 
\end{align*}
In practice, just like we can't shoot neutrinos through an anti-Earth, it is not reasonable to assume we can build both neutrino sources and detectors at two different positions on the Earth's surface, so we are stuck with an improper test of T invariance, where we compare $P^{\rm matter}_{\mu@a\to e@b}$ and $P^{\rm matter}_{e@a\to \mu@b}$. Improper T-invariance tests, however, are qualitatively different from improper CP-invariance tests. If the  $a\to b$ trajectory is indistinguishable from the $b\to a$ trajectory,  $P^{\rm matter}_{\mu@a\to e@b} = P^{\rm matter}_{\mu@b\to e@a}$ and the proper and improper tests of T invariance are identical. As far as neutrino oscillations are concerned,  the  $a\to b$ trajectory is indistinguishable from the $b\to a$ trajectory if the matter potential is symmetric within the $a\leftrightarrow b$ segment. In summary, matter-induced T-invariance violation (TV) is proportional to the asymmetry of the matter potential \cite{Kuo:1987km,Akhmedov:2001kd,Miura:2001pia,Ohlsson:2003nb,Akhmedov:1999uz}.

Here, we explore T invariance in neutrino oscillations, concentrating on the nontrivial role of matter effects. We aim at a pedagogical discussion and will repeat some well-known results for the sake of completeness. We restrict our discussion to the quantum-mechanical formalism of neutrino oscillations, as made clear in Section~\ref{sec:formalism}. We discuss the two-flavor case in some detail in Section~\ref{sec:2f}. The two-flavor case is special in the sense that it requires \emph{improper} T invariance to be exact under all circumstances. We discuss the three-flavor case in Section~\ref{sec:3f}. In Section~\ref{sec:3f}, we also discuss some of the practical implications of matter effects in possible future experimental setups and compare the effects of matter in CP-violating, T-violating, and CPT-violating observables. We summarize our findings and present some concluding remarks in Section~\ref{sec:conclusion}. 

\section{Formalism}
\label{sec:formalism}
\setcounter{equation}{0}

We assume all neutrino sources and detectors of interest are such that we can treat neutrinos as ultra-relativistic coherent linear superpositions of the different neutrino mass eigenstates $\nu_i$ with masses $m_i$ for $i=1,2,3,\ldots,N$ and a common, well-defined energy $E$. We will be interested in circumstances where the neutrinos are produced and detected as flavor eigenstates. As usual, the weak-interaction or flavor states are related to the mass eigenstates via
\begin{equation}
|\nu_{\alpha}\rangle = U_{\alpha i}^*|\nu_i\rangle,
\end{equation} 
where $\alpha=e,\mu,\tau,\ldots$, and $U_{\alpha i}$ are the elements of a unitary matrix. When $N\ge 4$, we assume there are $N-3$ ``sterile neutrinos'' along with the three weak-interaction eigenstates but, unless otherwise noted, we are not interested in $N\ge 4$. Both the mass eigenstates and the weak eigenstates form a complete basis for the neutrino flavor wave function. 

 Antineutrino flavor eigenstates are also related to the antineutrino mass eigenstates via 
\begin{equation}
|\bar{\nu}_{\alpha}\rangle = U_{\alpha i}|\bar{\nu}_i\rangle.
\end{equation} 
CPT invariance ensures that the neutrino and antineutrino mass eigenstates have identical masses and that the same $U_{\alpha i}$ appears in the two equations above (modulo the complex conjugation).

Flavor evolution along the  the source--detector trajectory (coordinate $x$, baseline $L$) is governed by the following Schr\"odinger-like equation:
\begin{equation}
i\frac{\rm d}{{\rm d}x} |\nu\rangle = H|\nu\rangle,
\label{eq:Sch}
\end{equation}
where $H$ (the Hamiltonian) is a $3\times 3$ Hermitian matrix. Assuming no new interactions other than those of the standard model, we can write, for neutrinos,
\begin{equation}
H = \sum_i\frac{m_i^2}{2E}|\nu_i\rangle\langle\nu_i| + A(x)|\nu_e\rangle\langle\nu_e|,
\label{eq:Hnu}
\end{equation}
where $A=\sqrt{2}G_Fn_e(x)$,  $G_F$ is the Fermi constant, and $n_e(x)$ is the (in general, position dependent) electron-number density of the medium, which we assume is neutral. If there are other vector-mediated interactions between neutrinos and ordinary matter, their effects on the unitary flavor evolution can be included by modifying the second term in Eq.~(\ref{eq:Hnu}), the matter potential, to $\sum_{\alpha\beta}A_{\alpha\beta}(x)|\nu_\alpha\rangle\langle\nu_\beta|$ where $A_{\alpha\beta}(x)$ will depend on the details of the new interaction and the relevant local properties of the medium (e.g., electron or neutron number densities).

Antineutrinos also obey Eq.~(\ref{eq:Sch}), with a different Hamiltonian. Assuming no new interactions other than those of the standard model,
\begin{equation}
H = \sum_i\frac{m_i^2}{2E}|\bar{\nu}_i\rangle\langle\bar{\nu}_i| - A(x)|\bar{\nu}_e\rangle\langle\bar{\nu}_e|.
\label{eq:Hnubar}
\end{equation}
It is clear that, as long as $n_e\neq 0$, neutrinos and antineutrinos will oscillate differently; they have different Hamiltonians. 

The probability that a neutrino\footnote{For the remainder of this section, we restrict the discussion to neutrino states (as opposed to antineutrinos).} is produced as a $\nu_{\alpha}$ at point $a$ characterized by $x=0$ and detected as a $\nu_{\beta}$ at point $b$, characterized by $x=L$ ($\alpha,\beta=e,$ $\mu,$ $\tau$), can be written as  $P_{\alpha\beta}=|V_{\beta\alpha}(L)|^2$, where $V$ is the unitary flavor-evolution matrix:
\begin{equation}
\langle\nu_{\beta}|\nu(L)\rangle = \langle\nu_{\beta}|V(L)|\nu_{\alpha}\rangle = V_{\beta\alpha}(L).
\end{equation}
T invariance is probed by comparing $P_{\alpha\beta}$ with $P_{\beta\alpha}$ or, equivalently, $|V_{\beta\alpha}(L)|^2$ with $|V_{\alpha\beta}(L)|^2$. For constant matter, $V(L)=e^{-iHL}$ and 
\begin{equation}
V^t = \left[e^{-iHL}\right]^t = e^{-iH^tL} = e^{-iH^*L},
\end{equation}
where $t$ indicates the transpose and we took advantage of the fact that $H$ is Hermitian. If $H$ is real in the flavor basis, for constant matter, it trivial to see that $V=V^t$: $V$ is a symmetric matrix and, trivially, $|V_{\alpha\beta}|^2=|V_{\beta\alpha}|^2$. In the absence of new interactions, $H$, in the flavor basis, is not real only if some of the $U_{\alpha i}$ are also not real. When there are new interactions, $H$ is real if all new physics couplings and all weak couplings -- captured by the mixing matrix $U$ -- can be chosen real. When $H$ is real in the flavor basis, we will state that there is no \textit{intrinsic} CP-invariance violation or TV. 

Next, we consider $A(x)$ that is piecewise constant: $A(x) = A_n$ for $x\in[x_{n-1},x_n]$, $n=1,2,\ldots, N$, where $x_0=0$, $x_{N}=L$. All matter potentials of interest can be treated as piecewise constant. Defining $H_n=H(A_n)$ and $V_n=e^{-iH_n(x_n-x_{n-1})}$, 
\begin{equation}
V^{ab} = V_NV_{N-1}\cdots V_2V_1,
\label{eq:V}
\end{equation}
describes the flavor evolution of neutrinos produced at $x=0$ (point $a$) and detected at $x=L$ (point $b$); $P^{ab}_{\alpha\beta}=|V^{ab}_{\beta\alpha}|^2$. 

Similarly, the flavor evolution of a neutrino produced at $x=L$ (point $b$) and later detected at $x=0$ (point $a$) is characterized by 
 \begin{equation}
V^{ba} = V_1V_{2}\cdots V_{N-1}V_N,
\label{eq:Vback}
\end{equation} 
and $P^{ba}_{\alpha\beta}=|V^{ba}_{\beta\alpha}|^2$. Note that the $x_n-x_{n-1}$ factors in the $V_n$ position-evolution matrices are all positive, i.e., we don't consider negative propagation distances. One way to derive  Eq.~(\ref{eq:Vback}) is to relabel the $x_n\to L-x_n$ and follow the same logic that led to Eq.~(\ref{eq:V}).

If $H$ is real in the flavor basis, $V_n$ are symmetric for all $n$, as argued above. For a piecewise constant $A$, however, there is no reason for $V$ to be symmetric -- $V^t = V^t_1V_{2}\cdots V^t_{N-1}V^t_N\neq V$ even if all $V_n=V_n^t$. Hence, matter-induced, \emph{improper} TV is generally expected: $P^{ab}_{\alpha\beta}\neq P^{ab}_{\beta\alpha}$.\footnote{As we discuss in the next section, the case of two flavors is special.} This is similar to the well-known case of matter-induced CP-invariance violation. Since the matter potential is not CP-invariant, one expects $P_{\alpha\beta}\neq P_{\bar{\alpha}\bar{\beta}}$ even when there is no intrinsic CP-invariance violation; and since the matter potential is not, in general, T-invariant, $P^{ab}_{\alpha\beta}\neq P^{ab}_{\beta\alpha}$ even when there is no intrinsic TV. 

A \emph{proper} test of T-invariance conservation, as discussed in the introduction, is the comparison of $P^{ab}_{\alpha\beta}$ and $P^{ba}_{\beta\alpha}$. For that we need
\begin{equation}
(V^{ba})^t = V^t_NV^t_{N-1}\cdots V^t_{2}V^t_1.
\end{equation}
When $V_n=V_n^t$, $V^{ba}_{\alpha\beta}=V^{ab}_{\beta\alpha}$ and $P^{ba}_{\alpha\beta}=P^{ab}_{\beta\alpha}$. This is as expected. $P^{ab}_{\alpha\beta}$ and $P^{ba}_{\beta\alpha}$ are related by the T-conjugation operation and when there is no intrinsic T-invariance violation -- e.g., when the weak neutrino--charged-lepton couplings, encoded in the $U_{\alpha i}$, are real --  $P^{ab}_{\alpha\beta}=P^{ba}_{\beta\alpha}$. 

Finally, symmetric matter potentials -- $A(x) = A(L-x)$ and $V_n = V_{N+1-n}$ -- are special. In this case, 
\begin{equation}
V^{ab} = V_1V_2\cdots V_2V_1 = V^{ba},
\end{equation}
and proper tests of T invariance are identical to improper tests of T invariance. This implies, for example, that for any symmetric matter potential, if there is no intrinsic TV, $P^{ab}_{\alpha\beta}=P^{ab}_{\beta\alpha}$; matter-induced T-invariance violating effects require asymmetric potentials. 

\subsection{Mass Basis}
\setcounter{footnote}{0}

We argued that if the Hamiltonian is real in the flavor basis (equivalent to real $U$, and assuming no interactions beyond those in the SM) and the matter potential is symmetric, then T invariance holds: $P_{\alpha\beta}=P_{\beta\alpha}$. On the other hand, for an asymetric matter potential, even if there is no intrinsic TV,  matter-induced, improper TV is possible. In this subsection, we revisit some of these issues, this time in the mass basis.  

In the mass basis, assuming no new interactions, the Hamiltonian -- Eq.~(\ref{eq:Hnu}) --  is explicitly real as long as we pick a parameterization for $U$ where $U_{ei}$ are real $\forall i=1,2,3$.\footnote{$A(x)|\nu_e\rangle\langle\nu_e|=A(x)\sum_{i,j}U_{ei}U_{ej}^*|\nu_i\rangle\langle\nu_j|$, and $A(x)$ is real.} Here, it is convenient to pick the standard PDG parameterization, where $U_{e3}=\sin\theta_{13}e^{-i\delta_{\rm CP}}$, multiplied on the right by diag$(1,1,e^{+i\delta_{\rm CP}})$, i.e.,
{\small
\begin{equation}
U=\left(\begin{array}{ccc} \cos\theta_{12}\cos\theta_{13} & \sin\theta_{12}\cos\theta_{13} & \sin\theta_{13} \\  
-\sin\theta_{12}\cos\theta_{23}-\cos\theta_{12}\sin\theta_{23}\sin\theta_{13}e^{i\delta_{\rm CP}}  & \cos\theta_{12}\cos\theta_{23}-\sin\theta_{12}\sin\theta_{23}\sin\theta_{13}e^{i\delta_{\rm CP}}  & \cos\theta_{13}\sin\theta_{23}e^{i\delta_{\rm CP}} \\ 
\sin\theta_{12}\sin\theta_{23}-\cos\theta_{12}\cos\theta_{23}\sin\theta_{13}e^{i\delta_{\rm CP}}  & -\cos\theta_{12}\sin\theta_{23}-\sin\theta_{12}\cos\theta_{23}\sin\theta_{13}e^{i\delta_{\rm CP}}  & \cos\theta_{13}\cos\theta_{23}e^{i\delta_{\rm CP}} \end{array} \\ \right).
\label{eq:U}
\end{equation}}

The probability that a $\nu_i$ produced at $x=0$ would be detected as a $\nu_j$ at $x=L$ is $P_{ij}=|{\cal V}_{ji}|^2$ where ${\cal V}_{ij}$ are the elements of the flavor-evolution matrix in the mass basis. In vacuum, $P_{ij}=\delta_{ij}$; the neutrino mass eigenstates are eigenstates of the propagation Hamiltonian in vacuum. The situation in matter is less trivial. Nonetheless, taking advantage of the results derived earlier, since the Hamiltonian is real, in the absence of new interactions, $P_{ij}=P_{ji}$ as long as the matter potential is symmetric. This does not depend on whether there is intrinsic TV, i.e., if $\delta_{\rm CP}\neq 0, \pi$. More generally, $P_{ij}$ does not depend on $\theta_{23}$ or $\delta_{\rm CP}$ at all simply because the Hamiltonian in the mass basis does not depend on $\theta_{23}$ or $\delta_{\rm CP}$. This is also true of the amplitudes: ${\cal V}_{ij}$ are functions of $\theta_{12},\theta_{13}$, the mass-squared differences, $A(x)$ and $L$. This statement does not depend on the symmetry of the matter potential; $P_{ij}$ does not depend on $\theta_{23}$ or $\delta_{\rm CP}$ for any distribution of ordinary matter.

While it is difficult, in practice, to study the oscillation of mass eigenstates, the statements above have consequences for flavor oscillations. ${\cal V}_{ij}$ to $V_{\alpha\beta}$ are related: 
\begin{equation}
V_{\alpha\beta} = {\cal V}_{ij} U_{\alpha i}U^*_{\beta j}.
\end{equation}
In this form, it easy to assess that $V_{\alpha\beta}$ can depend both on $\theta_{23}$ and $\delta_{\rm CP}$ but that the dependency is exclusively encoded in the elements of $U$. Using Eq.~(\ref{eq:U}) and the fact that ${\cal V}$ does not depend on $\delta_{\rm CP}$, one can write,
\begin{equation}
V_{\alpha e} = a_{\alpha} + b_{\alpha}e^{i\delta_{\rm CP}},
\end{equation}
where $a_{\alpha},b_{\alpha}$ are complex but do not depend on $\delta_{CP}$. Furthermore, for $\alpha = e$, it is trivial to check that $b=0$ so $P_{ee}$ never depends on $\delta_{\rm CP}$ (or $\theta_{23}$) \cite{Kuo:1989qe}. On the other hand, for $\alpha=\mu,\tau$, 
\begin{equation}
P_{e\alpha} = |V_{\alpha e}|^2 =A_{\alpha} + B_{\alpha}\cos\delta_{\rm CP} + C_{\alpha}\sin\delta_{\rm CP},
\label{eq:Pdelta}
\end{equation}
where $A_{\alpha},B_{\alpha},C_{\alpha}$ are real and do not depend on $\delta_{\rm CP}$. This remarkable result was first demonstrated in \cite{Kimura:2002hb}. It was also discussed in \cite{Akhmedov:2001kd,Miura:2001pia,Lipari:2001ds,Ohlsson:2003nb}. Here, we highlight that the result also holds for any matter distribution, as long as there are no new interactions. Furthermore, using $\sum_{\alpha} P_{e\alpha}=1$, one can show that $B_{\mu}=-B_{\tau}$ and $C_{\mu}=-C_{\tau}$.

Equivalently, 
\begin{equation}
P_{\alpha e} = |V_{e\alpha}|^2 =A'_{\alpha} + B'_{\alpha}\cos\delta_{\rm CP} - C'_{\alpha}\sin\delta_{\rm CP},
\label{eq:Pdelta2}
\end{equation}
In general, $A'_{\alpha},B'_{\alpha},C'_{\alpha}$ are different from $A_{\alpha},B_{\alpha},C_{\alpha}$.\footnote{We defined the coefficients in such a way that $A'_{\alpha}=A_{\alpha},B'_{\alpha}=B_{\alpha},C'_{\alpha}=C_{\alpha}$ in vacuum or for a symmetric matter potential.}  We will use Eq.~(\ref{eq:Pdelta}) when discussing bi-probability plots in Sec.~\ref{sec:3f}.

One may wonder whether the $\delta_{\rm CP}$ dependency of $P_{\mu\mu},\ P_{\mu\tau},\ P_{\tau\mu},\ P_{\tau\tau}$ is the same as the one of Eq.~(\ref{eq:Pdelta}), i.e., whether these only depend linearly on $\cos\delta_{\rm CP}$ and $\sin\delta_{\rm CP}$. Here, we quickly demonstrate that $P_{\mu\mu}$ also depends on terms proportional to $\cos^2\delta_{\rm CP}$. Here we compute the oscillation in vacuum. Generalizing our conclusions to oscillation in matter is straightforward. 

In vacuum, we may choose to write $P_{\mu\mu}$ purely as a function of $|U_{\mu 2}|^2$, $|U_{\mu 3}|^2$ (by unitarity), and functions of $\Delta m^2_{21}$ and $\Delta m^2_{31}$. The only $\delta_{\rm CP}$ dependency in these comes in $|U_{\mu 2}|^2$, and so the only terms in $P_{\mu\mu}$ that do \textit{not} follow $A + B\cos\delta_{\rm CP} + C\sin\delta_{\rm CP}$ must arise in terms proportional to $|U_{\mu 2}|^2$. One term arising in the probability is
\begin{align}
P_{\mu\mu} \supset -4\left\lvert U_{\mu 2}\right\vert^2 \left(1 - 4\left\lvert U_{\mu 2}\right\rvert^2\right) \sin^2\left(\frac{\Delta m_{21}^2 L}{4E}\right).
\end{align}
In the $|U_{\mu 2}|^4$ piece, a term proportional to $\cos^2\delta_{\rm CP}$ appears. It takes the form
\begin{align}
P_{\mu\mu} \supset \sin^2\left(2\theta_{12}\right) \sin^2\left(2\theta_{13}\right) \sin^2\theta_{13} \cos^2\delta_{\rm CP} \sin^2\left(\frac{\Delta m_{21}^2 L}{4E}\right).
\end{align}
Using unitarity constraints, it follows immediately that  $P_{\mu\mu},\ P_{\mu\tau},\ P_{\tau\mu},\ P_{\tau\tau}$ all depend on $\cos^2\delta_{\rm CP}$. Using the formalism developed above, it is easy to extend this result to $P_{\mu\mu},\ P_{\mu\tau},\ P_{\tau\mu},\ P_{\tau\tau}$ in matter. 

At first, this lack of ``symmetry'' when it comes to the dependency of all $P_{\alpha\beta}$ on $\delta_{CP}$ may appear surprising. However, upon some reflection, it is easy to see that the definition of $\delta_{CP}$ is not flavor-indifferent.  The traditional parameterization of the leptonic mixing matrix (and Eq.~(\ref{eq:U})), in some sense, singles out both the electron row and the $\nu_3$ column; both of them are designed to only depend on two of the three mixing angles and the electron row can be chosen real.

\section{Two Flavors}
\label{sec:2f}
\setcounter{equation}{0}

If there are only two neutrino species -- for concreteness, $\nu_e$ and $\nu_{\mu}$ as linear superpositions of $\nu_1$ and $\nu_2$ -- the situation is qualitatively different. For all $2\times 2$ Hamiltonians (assuming they are Hermitian), 
\begin{equation}
V(L) = \left(\begin{array}{cc} V_{ee} & V_{\mu e} \\ V_{e\mu} & V_{\mu\mu}\end{array}\right),
\end{equation}
and $P_{\alpha\beta}=|V_{\beta\alpha}|^2$. $V_{\alpha\beta}$ are functions of $L$ and the Hamiltonian parameters. Unitary evolution, $VV^{\dagger}=\mathbb{1}$,  translates into 
\begin{equation}
P_{ee} + P_{e\mu}=1, ~~~~~P_{\mu e} + P_{\mu\mu}=1.
\label{eq:conserved1}
\end{equation}
This can be interpreted as the conservation of probability: the probability that neutrino produced as a $\nu_e$ is later measured as a $\nu_e$ or a $\nu_{\mu}$ is 100\% for all $L$ and all Hamiltonians. These equations, of course, generalize for more neutrino species, $\sum_{\beta}P_{\alpha\beta}=1$, $\forall \alpha$ . Similarly,  $V^{\dagger}V=\mathbb{1}$ translates into
\begin{equation}
P_{ee} + P_{\mu e}=1, ~~~~~P_{e\mu} + P_{\mu\mu}=1.
\label{eq:conserved2}
\end{equation}
One characteristic implication of Eqs.~(\ref{eq:conserved2}), and their generalization for more neutrinos species -- $\sum_{\alpha}P_{\alpha\beta}=1$, $\forall \beta$ --  is that if equal numbers of all neutrino flavors are produced, the probabilities of detecting neutrinos with a given flavor are always identical. The four equations in Eqs.~(\ref{eq:conserved1}, \ref{eq:conserved2}) allow one to relate all four $P_{\alpha\beta}$: $P_{ee}=P_{\mu\mu}=1-P_{e\mu}=1-P_{\mu e}$. In particular, $P_{e\mu}=P_{\mu e}$ for all $L$ and all Hamiltonians. In summary, if there are only two neutrino flavors, there is no improper TV in neutrino oscillations, independent from the presence of intrinsic TV in the theory. 

\begin{figure}[htbp]
\begin{center}
\includegraphics[width=0.45\linewidth]{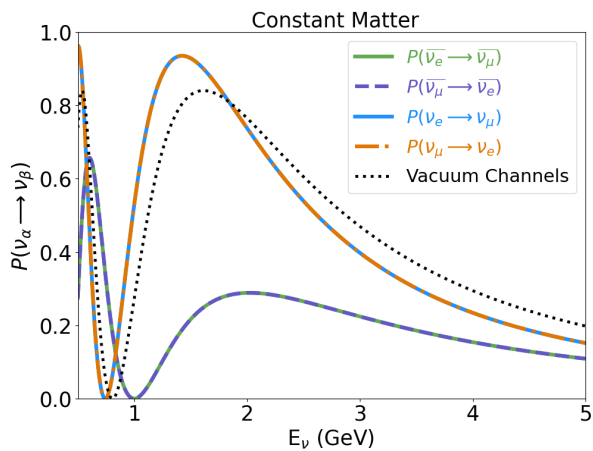}
\caption{Neutrino and antineutrino oscillation probabilities, assuming there are only two flavors, $\nu_e$ and $\nu_{\mu}$. The oscillation parameters are $\Delta m^2=10^{-3}$~eV$^2$, $|U_{e2}|^2=|U_{\mu1}|^2=\sin^2\theta=0.3$ and we define $m_1^2<m_2^2$. $L=2000$~km. The dotted lines correspond to oscillations in vacuum while the solid and dashed lines correspond to oscillations in constant matter with $A=1.1\times 10^{-3}$~eV$^2$/GeV.}
\label{fig:2flavors_constant}
\end{center}
\end{figure}
We briefly present and discuss a few concrete examples. Fig.~\ref{fig:2flavors_constant} depicts several oscillation probabilities as a function of the neutrino energy assuming a neutrino mass-squared difference $\Delta m^2=10^{-3}$~eV$^2$, a mixing parameter $|U_{e2}|^2=|U_{\mu1}|^2=\sin^2\theta=0.3$ and $L=2000$~km. We define $m_1^2<m_2^2$. The dotted lines correspond to vacuum oscillations of both neutrinos and antineutrinos; the four vacuum oscillation probabilities are identical and the curves lie on top of one another. The solid and dashed curves correspond to the oscillation probabilities for neutrinos and antineutrinos, as labelled in the figure, assuming the (anti)neutrinos traverse ordinary matter with constant  $A=1.1\times 10^{-3}$~eV$^2$/GeV, which corresponds to constant $\rho = 5.7$ g/cm$^3$ isoscalar matter.\footnote{Throughout, we will quote matter potentials in units of eV$^2$/GeV. For isoscalar matter with density $\rho$, $$\frac{A}{5.0\times 10^{-4}~\rm{eV^2/GeV}}=\frac{\rho}{2.6~\rm{g/cm^3}}.$$} There are strong matter-induced CP-invariance violating and CPT-invariance violating effects ($P_{\alpha\beta}\neq P_{\bar{\alpha}\bar{\beta}}$, $P_{\alpha\beta}\neq P_{\bar{\beta}\bar{\alpha}}$, respectively) but T invariance is observed ($P_{\alpha\beta}= P_{\beta\alpha}$, $P_{\bar{\alpha}\bar{\beta}}= P_{\bar{\beta}\bar{\alpha}}$).

\begin{figure}[htbp]
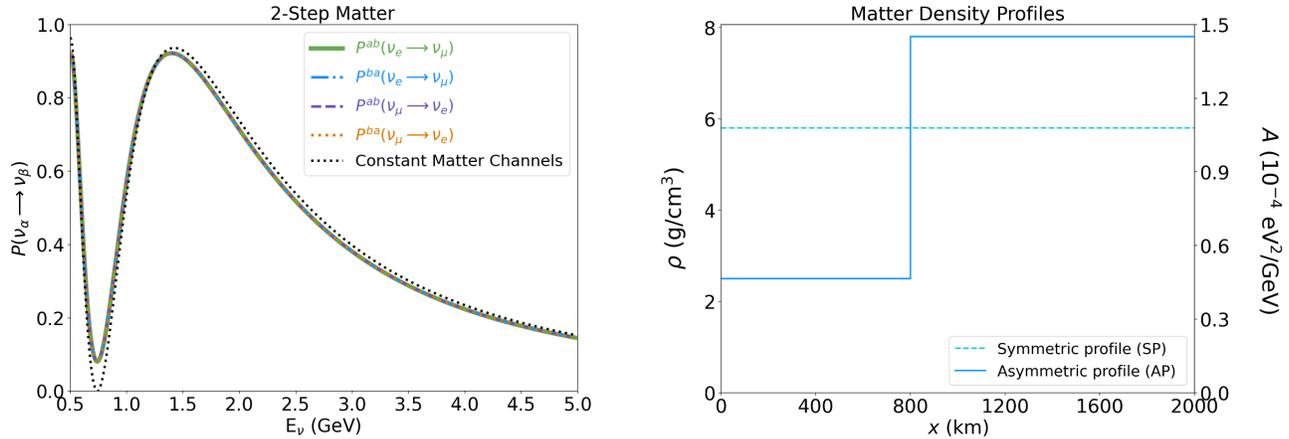

\begin{center}
\includegraphics[width=0.45\linewidth]{SM_Aym_matter_2F_La800km_Aa5p0times10toMinus4_Lb1200km_Ab1p5times10toMinus3_SineSqTheta0pt3_deltaMsq10toMinus3_energy0pt5to5GeV_NO.pdf}
\includegraphics[width=0.45\linewidth]{MDPs.pdf}
\caption{Left: Neutrino and antineutrino oscillation probabilities, assuming there are only two flavors, $\nu_e$ and $\nu_{\mu}$. The oscillation parameters are $\Delta m^2=10^{-3}$~eV$^2$, $|U_{e2}|^2=|U_{\mu1}|^2=\sin^2\theta=0.3$ and we define $m_1^2<m_2^2$. $L=2000$~km. The black dotted lines (all identical) correspond to oscillations in constant matter with $A=1.1\times 10^{-3}$~eV$^2$/GeV while the colorful lines correspond to oscillations in piecewise constant matter with $A_1=5\times 10^{-4}$~eV$^2$/GeV, $x_1=800$~km and $A_2=1.5\times 10^{-3}$~eV$^2$/GeV, $x_2=800+1200$~km (right panel). The different oscillation channels are labelled in the plot. The $ab$ superscript indicates neutrinos produced at $a$ and detected at $b$ while $ba$ indicates the reversed process. See text for details.}
\label{fig:2flavors_asymmetric}
\end{center}
\end{figure}
Fig.~\ref{fig:2flavors_asymmetric}(left) depicts neutrino oscillation probabilities for an asymmetric two-step, piecewise constant matter potential, depicted in Fig.~\ref{fig:2flavors_asymmetric}(right). Using the notation defined in the last section, $A_1=5\times 10^{-4}$~eV$^2$/GeV, $x_1=800$~km and $A_2=1.5\times 10^{-3}$~eV$^2$/GeV, $x_2=L=800+1200=2000$~km. As in the constant matter case, the oscillation parameters are $\Delta m^2=10^{-3}$~eV$^2$, $|U_{e2}|^2=|U_{\mu1}|^2=\sin^2\theta=0.3$ and we define $m_1^2<m_2^2$. We consider both the case where the neutrinos are produced at $x=0$ and detected at $x=L$ (superscript $ab$) and the case where the neutrinos are produced at $x=L$ and detected at $x=0$ (superscript $ba$). In order to gauge the impact of the asymmetric matter potential, Fig.~\ref{fig:2flavors_asymmetric} also depicts neutrino oscillation probabilities for a constant matter potential $A=1.1\times 10^{-3}$~eV$^2$/GeV. The values of $A,A_1,A_2$ are chosen such that for both the constant and asymmetric matter potentials the average value of the potential is the same. Here, even though the matter potential is not T-invariant, all neutrino oscillation probabilities are the same, as discussed earlier. To summarize, $P^{ab}_{e\mu}=P^{ab}_{\mu e}$ because of unitarity while $P^{ab}_{e\mu}=P^{ba}_{\mu e}$ because of T-invariance conservation. 

\begin{figure}[htbp]
\begin{center}
\includegraphics[width=0.45\linewidth]{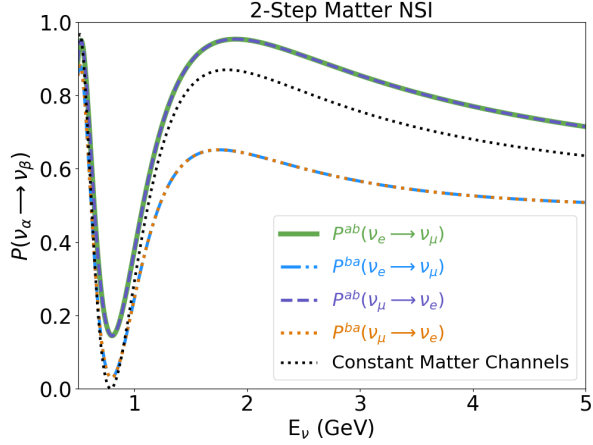}
\caption{Neutrino and antineutrino oscillation probabilities, assuming there are only two flavors, $\nu_e$ and $\nu_{\mu}$. The oscillation parameters are $\Delta m^2=10^{-3}$~eV$^2$, $|U_{e2}|^2=|U_{\mu1}|^2=\sin^2\theta=0.3$ and we define $m_1^2<m_2^2$. $L=2000$~km. Here, we assume the presence of non-standard neutrino--matter interactions that translate into Eq.~(\ref{eq:HNSI}). The black dotted lines (all identical) correspond to oscillations in constant matter with $A=1.1\times 10^{-3}$~eV$^2$/GeV while the colorful lines correspond to oscillations in piecewise constant matter with $A_1=5\times 10^{-4}$~eV$^2$/GeV, $x_1=800$~km and $A_2=1.5\times 10^{-3}$~eV$^2$/GeV, $x_2=800+1200$~km, depicted in Fig.~\ref{fig:2flavors_asymmetric}(right). The $ab$ superscript indicates neutrinos produced at $a$ and detected at $b$ while $ba$ indicates the reversed process. See text for details.}
\label{fig:2flavors_asymetric_TV}
\end{center}
\end{figure}

Finally, we postulate the existence of a new flavor-transforming neutrino--matter interaction that modifies the neutrino matter potential. Concretely, we assume the non-standard interactions translate into a modification of the neutrino matter potential in Eq.~(\ref{eq:Hnu}), which now includes
\begin{equation}
H_{\rm NSI} = i\frac{A(x)}{2}|\nu_e\rangle\langle\nu_\mu| - i\frac{A(x)}{2}|\nu_\mu\rangle\langle\nu_e|.
\label{eq:HNSI}
\end{equation}
The effects of the new interaction are both CP-odd and T-odd. Fig.~\ref{fig:2flavors_asymetric_TV} depicts neutrino oscillation probabilities for the two-step, piecewise constant matter potential depicted in Fig.~\ref{fig:2flavors_asymmetric}. The oscillation parameters are identical to the ones used to generate Fig.~\ref{fig:2flavors_asymmetric} except for the presence of the non-standard interaction. In agreement with the general discussion early in this section, $P^{ab}_{e\mu}=P^{ab}_{\mu e}$ due to the unitarity of the flavor evolution and the fact there are only two neutrino species. However, the Hamiltonian is not T-invariant and hence $P^{ab}_{e\mu}\neq P^{ba}_{\mu e}$. Both the non-standard, T-violating interaction and the asymmetric matter potential are required in order to observe proper TV. As highlighted earlier, as long as the flavor evolution is unitarity, improper TV is impossible when there are only two neutrino species. For comparison, the figure also depicts the oscillation probabilities assuming a constant matter density. As discussed, $P^{ab}_{\mu e}=P^{ab}_{e\mu}=P^{ba}_{\mu e}=P^{ba}_{e\mu}$ -- the $a\to b$ and $b\to a$ trajectories are identical for constant matter -- because there is no improper TV for two flavors, independent of the Hamiltonian.

\section{Three Flavors and Some Practical Results}
\label{sec:3f}
\setcounter{equation}{0}
\setcounter{footnote}{0}

With three or more flavors, the general discussions in Sec.~\ref{sec:formalism} apply. We present a few concrete examples in order to illustrate how matter effects can impact studies of TV. There are three known neutrino flavors and most of the three-neutrino oscillation parameters have been measured, some with excellent precision. We make use of the standard PDG parameterization for the $3\times 3$ mixing matrix and concentrate on the so-called normal neutrino mass ordering, $m_1^2<m_2^2<m_3^2$, unless otherwise noted. When computing oscillation probabilities, we use the best-fit values of the oscillation parameters reported by the NuFit collaboration (2024 results)~\cite{Esteban:2024eli}: sin$^2 \theta_{12}$ = 0.308, sin$^2 \theta_{23}$ = 0.470, sin$^2 \theta_{13}$ = 0.02215, $\Delta m^2_{21}$ = 7.49$\times$10$^{-5}$ eV$^2$, and  $\Delta m^2_{31}$ = 2.513$\times$10$^{-3}$ eV$^2$.\footnote{For the inverted mass-ordering, we also use the best fit results from NuFit: sin$^2 \theta_{12}$ = 0.308, sin$^2 \theta_{23}$ = 0.550, sin$^2 \theta_{13}$ = 0.02231, $\Delta m^2_{21}$ = 7.49$\times$10$^{-5}$ eV$^2$, and  $\Delta m^2_{31}$ = -2.409$\times$10$^{-3}$ eV$^2$.}

\begin{figure}[htbp]
\begin{center}
\includegraphics[width=0.49\linewidth]{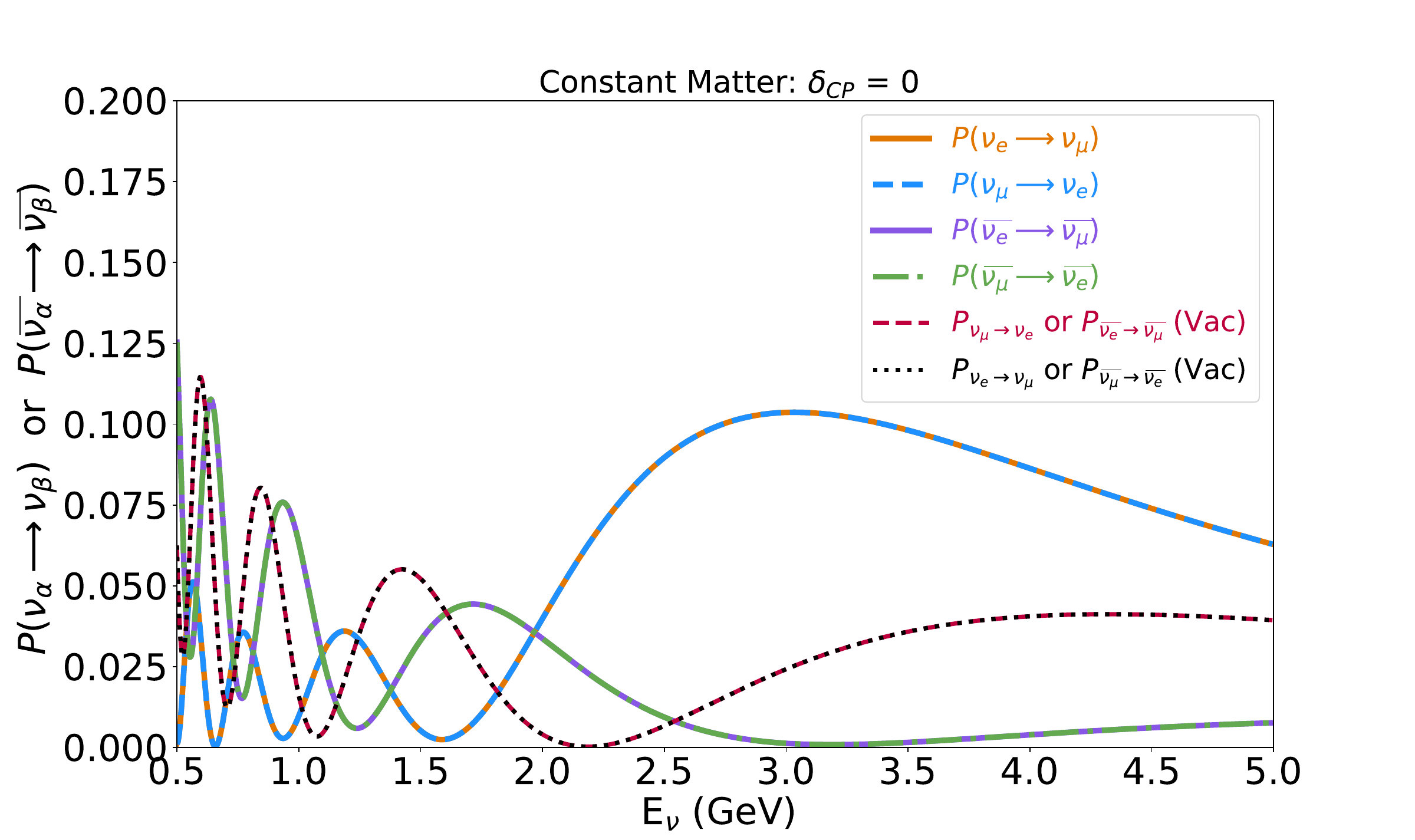}
\includegraphics[width=0.49\linewidth]{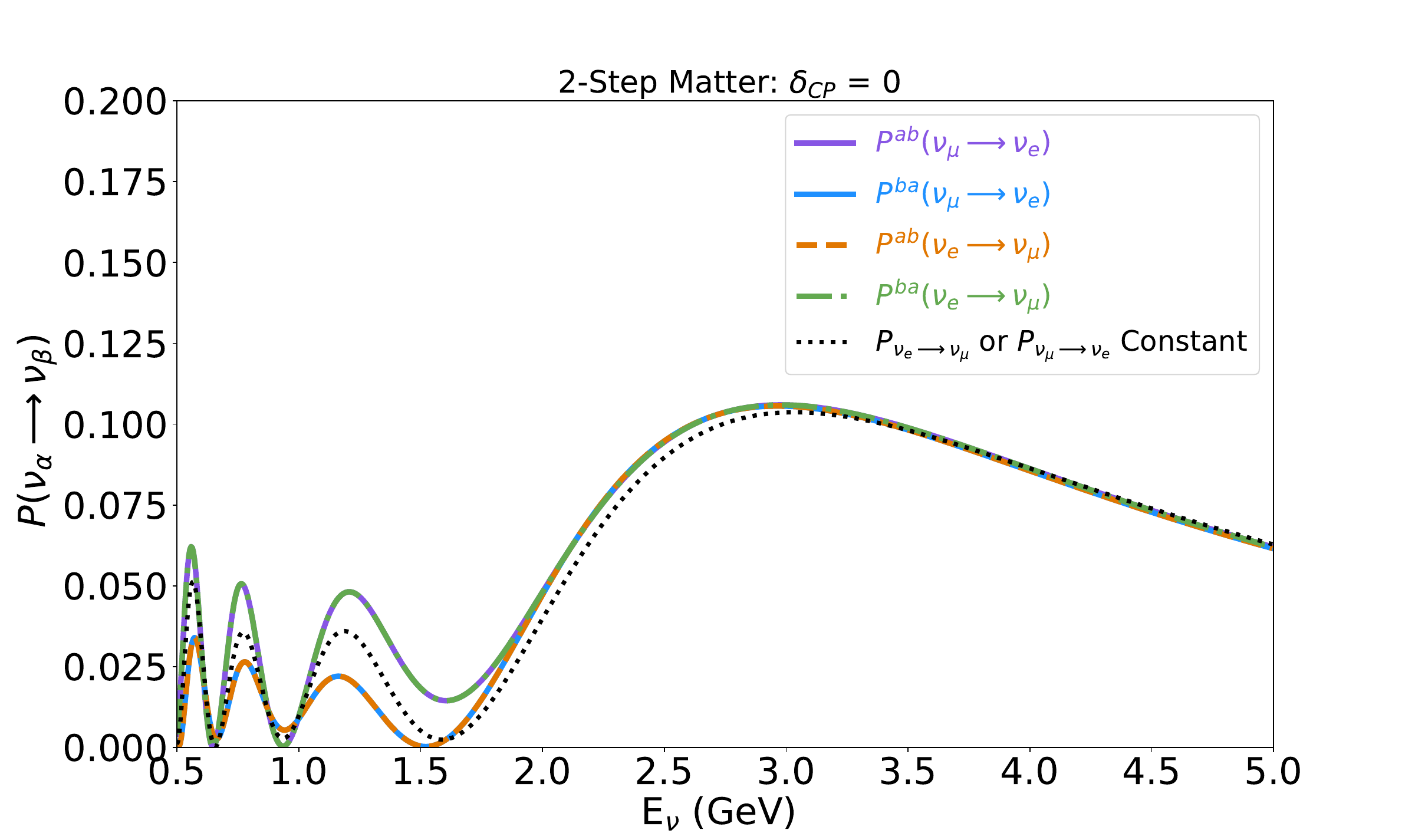}
\caption{Neutrino and antineutrino oscillation probabilities for $L=2000$~km and $\delta=0$. Left: Black dotted/red dashed lines, all identical, correspond to oscillations in vacuum while the colorful lines correspond to oscillations in constant matter with $A=1.1\times 10^{-3}$~eV$^2$/GeV.
Right: Black dotted lines correspond to oscillations in symmetric matter while the colored lines correspond to oscillations in piecewise constant matter with $A_1=5\times 10^{-4}$~eV$^2$/GeV, $x_1=800$~km and $A_2=1.5\times 10^{-3}$~eV$^2$/GeV, $x_2=800+1200$~km. The $ab$ superscript indicates neutrinos produced at $a$ and detected at $b$ while $ba$ indicates the reversed process. See text for details.
}
\label{fig:3flavors_symmetric_TC}
\end{center}
\end{figure}

Fig.~\ref{fig:3flavors_symmetric_TC}(left) depicts $P_{\mu e}$, $P_{e\mu}$, $P_{\bar{\mu}\bar{e}}$, and $P_{\bar{e}\bar{\mu}}$ as a function of the (anti)neutrino energy for $L=2000$~km and $\delta_{\rm CP}=0$ (i.e., T invariance is conserved in vacuum). The black dotted lines correspond to vacuum oscillations of both neutrinos and antineutrinos; the four vacuum oscillation probabilities are identical and the curves lie on top of one another. The colorful curves (solid and dashed) correspond to the oscillation probabilities for neutrinos and antineutrinos, as labelled in the figure, assuming they traverse ordinary matter with constant  $A=1.1\times 10^{-3}$~eV$^2$/GeV (equivalent to neutral matter with $\rho = 5.7$ g/cm$^3$). There are strong matter-induced CP-invariance violating and CPT-invariance violating effects ($P_{e\mu}\neq P_{\bar{e}\bar{\mu}}$, $P_{e\mu}\neq P_{\bar{\mu}\bar{e}}$, respectively) but T invariance is observed ($P_{e\mu}= P_{\mu e}$, $P_{\bar{e}\bar{\mu}}= P_{\bar{\mu}\bar{e}}$) . 

In contrast, we consider a non-constant matter potential in~\cref{fig:3flavors_symmetric_TC}(right). We still focus on $\delta_{\rm CP} = 0$, where T invariance is conserved in vacuum, depicting now just the neutrino oscillation probabilities $P_{\mu e}$ and $P_{e\mu}$. The non-constant matter potential is the same as the one considered in~\cref{sec:2f}: $A_1=5\times 10^{-4}$~eV$^2$/GeV, $x_1=800$~km and $A_2=1.5\times 10^{-3}$~eV$^2$/GeV, $x_2=L=800+1200=2000$~km (\cref{fig:2flavors_asymmetric} (right)). We consider both the case where the neutrinos are produced at $x=0$ and detected at $x=L$ (superscript $ab$) and the case where the neutrinos are produced at $x=L$ and detected at $x=0$ (superscript $ba$). In order to gauge the impact of the asymmetric character of the matter potential, Fig.~\ref{fig:3flavors_symmetric_TC}(right) also depicts neutrino oscillation probabilities for a constant matter potential with $A=1.1\times 10^{-3}$~eV$^2$/GeV. As a consequence of intrinsic T-invariance conservation, $P^{ab}_{e\mu}=P^{ba}_{\mu e}$. However, since the matter potential is asymmetric, there are matter-induced, improper T-violating effects:  $P^{ab}_{e\mu}\neq P^{ab}_{\mu e}$. These effects are largest at lower energies, when three-flavor effects are most visible -- at energies below the first oscillation maximum. For higher energies, ``solar'' effects are very small and the system can be described as if there were only two neutrino flavors. In this case, as discussed in Sec.~\ref{sec:2f}, $P^{ab}_{e\mu}$ and $P^{ab}_{\mu e}$ are guaranteed to be identical.

\begin{figure}[htbp]
\begin{center}
\includegraphics[width=0.49\linewidth]{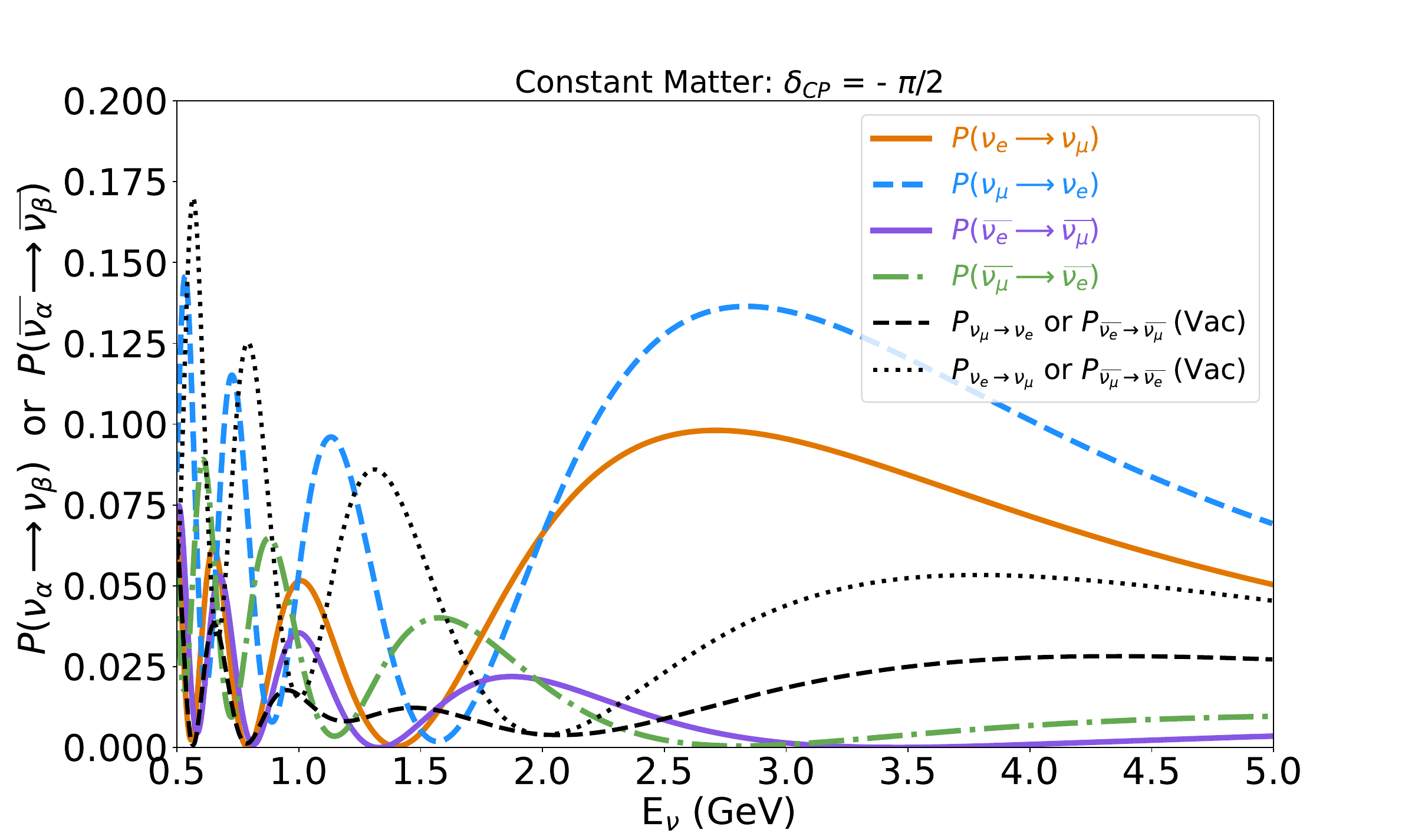}
\includegraphics[width=0.49\linewidth]{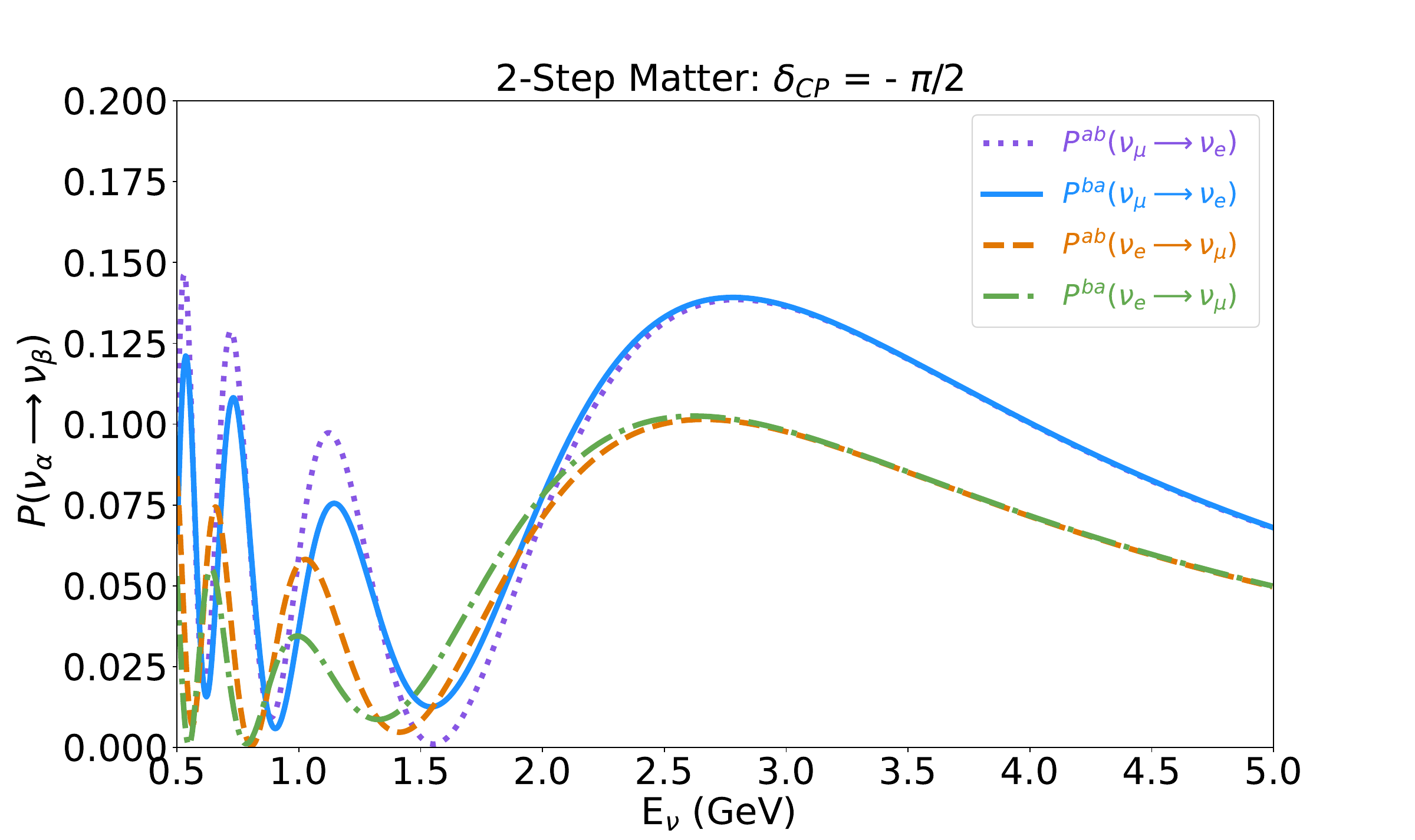}
\caption{Neutrino oscillation probabilities, $\delta_{\rm CP}=-\pi/2$. Left: The black dotted/dashed lines correspond to oscillations in vacuum -- note that CPT is conserved -- while the other four curves correspond to oscillations in constant matter with $A=1.1\times 10^{-3}$~eV$^2$/GeV. Right: Neutrino oscillations in piecewise constant matter with $A_1=5\times 10^{-4}$~eV$^2$/GeV, $x_1=800$~km and $A_2=1.5\times 10^{-3}$~eV$^2$/GeV, $x_2=800+1200$~km. The $ab$ superscript indicates neutrinos produced at $a$ and detected at $b$ while $ba$ indicates the reversed process. See text for details.}
\label{fig:3flavors_asymmetric_TV}
\end{center}
\end{figure}

Until now, we have focused on $\delta_{\rm CP}=0$ to enforce T-invariance conservation in vacuum and in constant (or symmetric) matter. We now explore the impact of matter effects, constant or not, when $\delta_{\rm CP} \neq 0$. This is depicted in~\cref{fig:3flavors_asymmetric_TV}; the left-hand panel focuses on constant, $A = 1.1\times 10^{-3}$~eV$^2$/GeV, matter, while the right one assumes the two-step potential described above and in~\cref{fig:2flavors_asymmetric} (right).~\cref{fig:3flavors_asymmetric_TV} reveals that the magnitude of the T-invariance violating effect (e.g., a T-odd asymmetry like $(P_{e\mu}-P_{\mu e})/(P_{e\mu}+P_{\mu e})$) is impacted by matter effects. 

In summary, for a symmetric matter potential, matter effects will not lead to matter-induced T-violating effects. If there is intrinsic TV (e.g., $\delta_{\rm CP}\neq 0,\pi$) matter effects will modify ``how much'' T invariance is violated. This is easy to see. In vacuum, 
\begin{equation}
P_{e\mu}-P_{\mu e} = 4\Im[U_{e1}U^*_{e2}U^*_{\mu 1}U_{\mu 2}]\left[\sin\left(\Delta_{21}L\right)-\sin\left(\Delta_{31}L\right)+\sin\left(\Delta_{32}L\right)\right],
\end{equation}
where $\Im[U_{e1}U^*_{e2}U^*_{\mu 1}U_{\mu 2}]$ is also known as $J$, the Jarlskog invariant, and $\Delta_{ij}=\Delta m^2_{ij}/2E$. In constant matter, one can choose to write  
\begin{equation}\label{eq:DeltaPMatter}
(P_{e\mu}-P_{\mu e})^{\rm matter} = 4\Im[\tilde{U}_{e1}\tilde{U}^*_{e2}\tilde{U}^*_{\mu 1}\tilde{U}_{\mu 2}]\left[\sin\left(\tilde{\Delta}_{21}L\right)-\sin\left(\tilde{\Delta}_{31}L\right)+\sin\left(\tilde{\Delta}_{32}L\right)\right],
\end{equation}
where $\tilde{U}_{\alpha i}$ are the elements of the matrix that diagonalizes the flavor-evolution Hamiltonian while $\tilde{\Delta}_{ij}=\lambda_i-\lambda_j$, where $\lambda_i$ are the eigenvalues of the flavor-evolution Hamiltonian. For a nonzero matter potential, $U_{\alpha i}\neq\tilde{U}_{\alpha i}$ and $\Delta_{ij}\neq \tilde{\Delta}_{ij}$ so $(P_{e\mu}-P_{\mu e})\neq (P_{e\mu}-P_{\mu e})^{\rm matter}$.\footnote{That this is correct can be argued, roughly, as follows. $\Delta_{ij}$ and $\tilde{\Delta}_{ij}$ have difference dependencies on the neutrino energy. Even if by happenstance $(P_{e\mu}-P_{\mu e})=(P_{e\mu}-P_{\mu e})^{\rm matter}$ for some value of the neutrino energy, the equality will not hold for all values of the neutrino energy.} However, when $J=0$, the ``matter'' Jarlskog invariant, $\tilde{J}=\Im[\tilde{U}_{e1}\tilde{U}^*_{e2}\tilde{U}^*_{\mu 1}\tilde{U}_{\mu 2}]$, also vanishes~\cite{Naumov:1991ju,Harrison:1999df}.

\begin{figure}[htbp]
\begin{center}
\includegraphics[width=\linewidth]{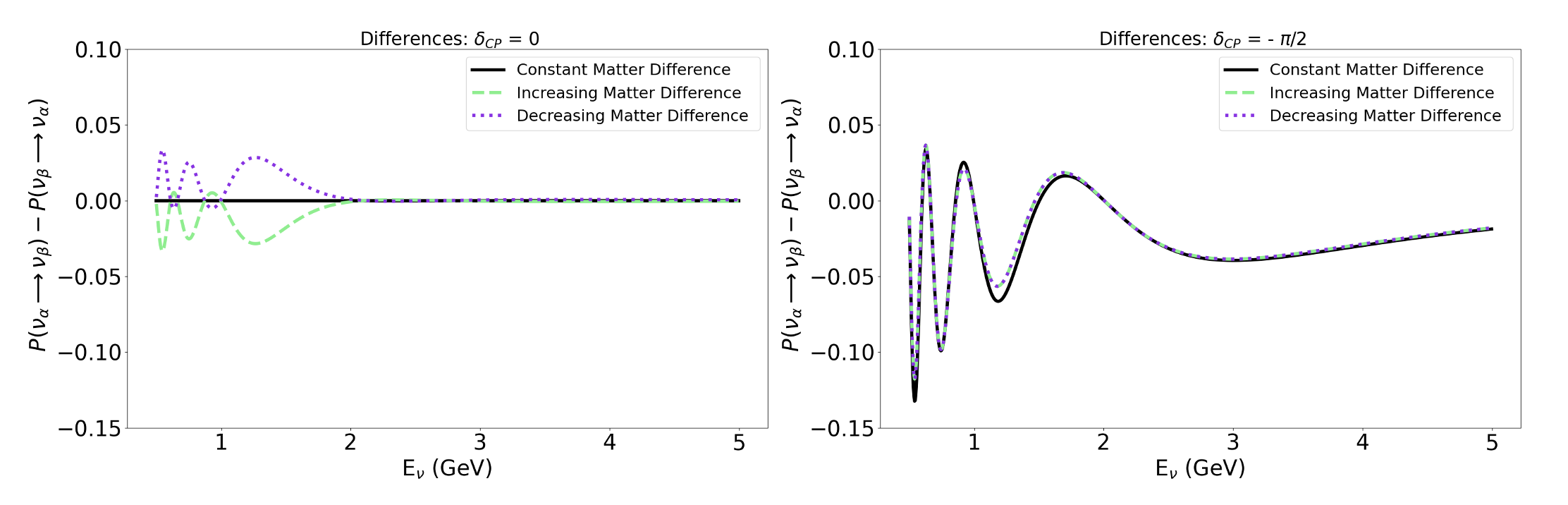}
\caption{$P_{\mu e}-P_{e\mu}$ for $\delta=0$ (left) and $\delta=-\pi/2$ (right). The black, solid curves correspond to constant matter with $A=1.1\times 10^{-3}$~eV$^2$/GeV. The green, dashed curves correspond to a piecewise constant matter with $A_1=5\times 10^{-4}$~eV$^2$/GeV, $x_1=800$~km and $A_2=1.5\times 10^{-3}$~eV$^2$/GeV, $x_2=800+1200$~km, while the purple dotted lines correspond to a piecewise constant matter with $A_1=1.5\times 10^{-3}$~eV$^2$/GeV, $x_1=1200$~km and $A_2=5\times 10^{-4}$~eV$^2$/GeV, $x_2=1200+800$~km. 
}
\label{fig:Diff}
\end{center}
\end{figure}
\cref{fig:Diff} depicts $P_{\mu e}-P_{e\mu}$ for $\delta_{\rm CP}=0$ (left) and $\delta_{\rm CP}=-\pi/2$ (right) and a constant and two different piecewise constant matter potentials: the one depicted in \cref{fig:2flavors_asymmetric} (right) [``increasing''], the other its ``mirror image,'' $x\to L-x$ [``decreasing'']. For $\delta_{\rm CP}=0$, we confirm that a constant matter potential does not induce TV while a piecewise constant matter potential does induce improper TV. The effect is equal opposite for the increasing relative to the decreasing matter potentials. The reason is that, for $\delta_{\rm CP}=0,$ there is no proper TV. The sum of the dashed and the dotted curves, taking into account that the two different potentials can be interpreted as the same potential where the production and detection points are reversed, is 
\begin{equation}
(P^{ab}_{\mu e} - P_{e\mu}^{ab}) + (P^{ba}_{\mu e} - P_{e\mu}^{ba}) = (P^{ab}_{\mu e} - P_{e\mu}^{ba}) - (P_{e\mu}^{ab} - P^{ba}_{\mu e}). 
\end{equation}
$P^{ab}_{\mu e} - P_{e\mu}^{ba}$ is the proper TV observable and the expression above therefore vanishes when there is no intrinsic TV (for any matter potential). Finally, For $\delta_{\rm CP}=-\pi/2$, the difference between the impact of symmetric and antisymmetric matter potentials is nonzero but quantitatively small. We explore this issue a little further in  the next two subsections.

\subsection{T-invariance Violation in Realistic Environments}
Throughout this section, we have explored the level of TV manifest in different situations, using the symmetric and asymmetric potentials of~\cref{fig:2flavors_asymmetric} (right) as a demonstrative case. The matter densities associated to these potentials  are relatively large compared to those relevant for long-baseline terrestrial experiments. The same is true of the total baseline length of $2000$~km. In this subsection, we briefly explore the level of TV in a more realistic scenario, focusing on the baseline length of the DUNE experiment ($1300$~km) as well as realistic matter density profiles. In particular, we use two matter density profiles corresponding to the Shen-Ritzwoller~\cite{Shen:2016kxw} and Crustal~\cite{Laske:2013xxx} profiles, analyzed in Refs.~\cite{Roe:2017zdw,Kelly:2018kmb}. While these two profiles are not symmetric, their asymmetries are small but appropriate for terrestrial long-baseline neutrino experiments.

\begin{figure}[!htbp]
\begin{center}
\includegraphics[width=\linewidth]{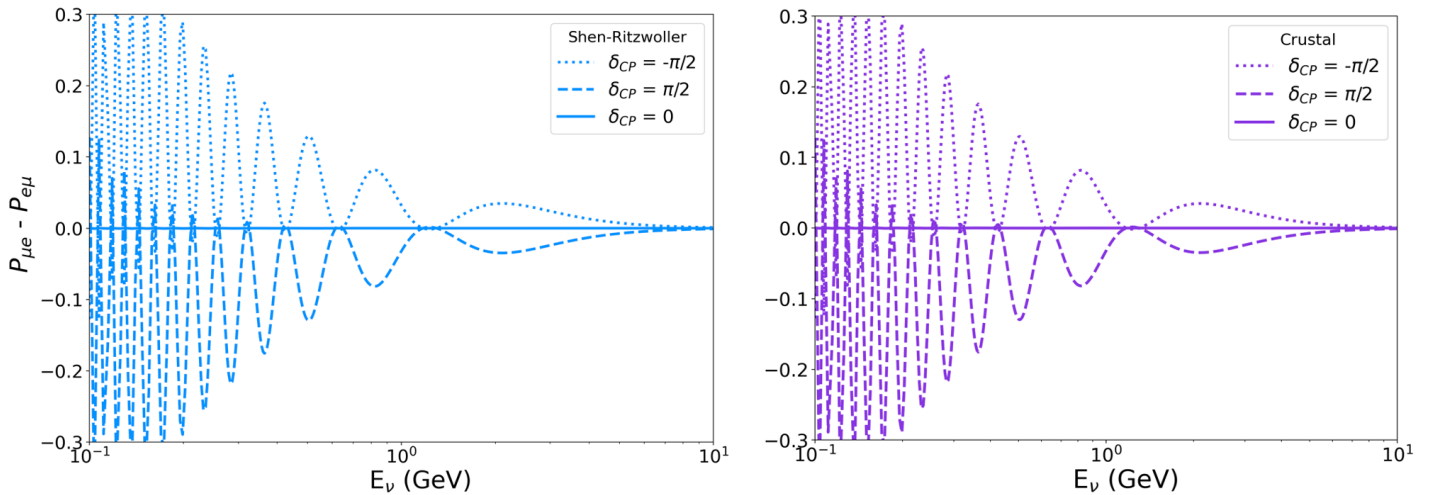}
\caption{Improper time-invariance violation present for DUNE baselines and energies ($L = 1300$~km) using the Shen-Ritzwoller~\cite{Shen:2016kxw} (Crustal~\cite{Laske:2013xxx}) matter density profile in the left (right) panel. For each panel, we display the difference of oscillation probabilities $P(\nu_\mu \to \nu_e) - P(\nu_e \to \nu_\mu)$ for three different choices of $\delta_{\rm CP}$: $0$~(solid), $+\pi/2$~(dashed), and $-\pi/2$~(dotted).\label{fig:DUNEDeltaP}}
\end{center}
\end{figure}

The oscillation-probability differences $P_{\mu e} - P_{e\mu}$ for these two matter density profiles are depicted in~\cref{fig:DUNEDeltaP}, where the left (right) panel corresponds to the Shen-Ritzwoller (Crustal) profile. For each panel, we consider $\delta_{\rm CP} = 0, \pm \pi/2$. While the two panels appear nearly identical, there are small differences between them. For $\delta_{\rm CP} = 0$,  $P_{\mu e} - P_{e\mu}$ is effectively zero for all energies, for both matter densiyy profiles. It is not, however,  \textit{exactly} zero as we will see in the coming discussion, but, for all practical purposes, the matter density profiles here are symmetric enough not to induce any meaningful TV if there is no intrinsic TV. 

While \cref{fig:DUNEDeltaP} reveals that $P_{\mu e} - P_{e\mu}$ may be observably large for $\delta_{\rm CP} = \pm \pi/2$, it bears distinguishing the impact of the \textit{size} of the matter density profile (with average densities of ${\sim}3$~g/cm$^3$) from that of the asymmetries in the profile over the $1300$~km of travel. To inspect this, for a given profile $\rho(x)$, we define an average matter density $\bar\rho$,
\begin{equation}
\bar\rho \equiv \frac{1}{L} \int_0^{L} \rho(x) dx.
\end{equation}
We then calculate the oscillation probabilities inside of the matter density $\rho(x)$ as well as in this averaged-matter, assuming a constant profile with uniform density $\bar\rho$. The asymmetry-induced TV, $\Delta P^{\rm assym.}$ is then defined by
\begin{equation}
\Delta P^{\rm assym.} \equiv \left(P_{\mu e} - P_{e\mu}\right) \rvert_{\rho(x)} - \left(P_{\mu e} - P_{e\mu}\right) \rvert_{\bar\rho}.
\end{equation}
By subtracting off the constant-matter-density difference, we isolate the amount of TV that is induced by the profile's asymmetry.

\begin{figure}[!htbp]
\begin{center}
\includegraphics[width=\linewidth]{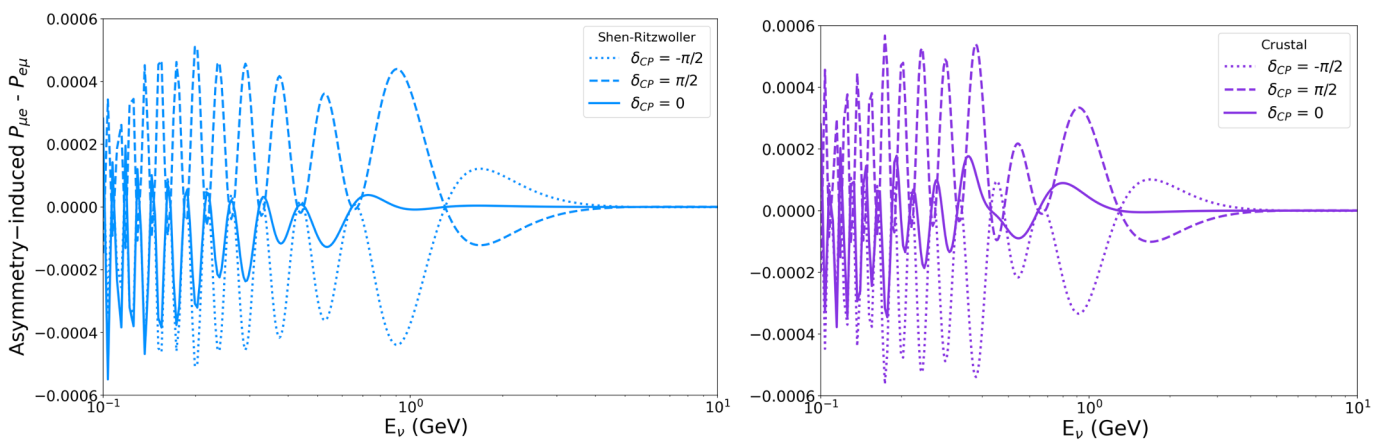}
\caption{The amount of asymmetry-induced T-invariance violation (see text for definition) at DUNE baseline/energies assuming two modestly asymmetric profiles: the Shen-Ritzwoller~\cite{Shen:2016kxw} (left) and the Crustal~\cite{Laske:2013xxx} (right). For each panel, three choices of $\delta_{\rm CP}$ are shown; other oscillation parameters are fixed to their best-fit values.\label{fig:AsymmetryInduced}}
\end{center}
\end{figure}

This quantity is depicted for the two matter density profiles in~\cref{fig:AsymmetryInduced} -- the Shen-Ritzwoller~\cite{Shen:2016kxw} profile on the left and the Crustal~\cite{Laske:2013xxx} on the right. For each panel, we present the asymmetry-induced violation for three choices of $\delta_{\rm CP}$: $0$ and $\pm \pi/2$. We see that for both profiles and for any choice of $\delta_{\rm CP}$ the asymmetry introduces very small amounts of matter-induced TV -- $\Delta P^{\rm assym.} \lesssim 6 \times 10^{-4}$ for all energies, where the oscillation probabilities themselves are at the level of ${\sim}5\%$ for these energies and baseline lengths -- an insignificant,  immeasurable amount.

Given the above discussion, we see that while matter may introduce relatively large induced TV (as long as the lepton sector has inherent CP violation), the amount of induced violation due to the matter density profile's asymmetry is limited to be very small -- at least when considering realistic amounts of asymmetry for terrestrial long-baseline accelerator experiments. In the next subsection, nonetheless, we futher explore the impact of these non-uniform matter density profiles. We will, at some point, make use of an exaggerated matter density profile, for illustrative purposes. 

\subsection{Bi-Probability Plots}\label{subsec:Biprobability}
A convenient way to visualize the amount of CP-invariance, T-invariance, and CPT-invariance violation and the impact (or lack thereof) of matter effects is to make use of bi-probability plots. These  have been utilized to understand many aspects of neutrino oscillations --  see, e.g., \cite{Kimura:2002hb,Minakata:2002qe,Blom:2004bk,Minakata:2001qm,Minakata_2010,Winter:2003ye,Ohlsson:2003nb,Singh:2016dpd,Minakata:2003jj}.  

As we discussed in detail in Sec.~\ref{sec:formalism} -- see Eqs.~(\ref{eq:Pdelta},\ref{eq:Pdelta2}) --
\begin{eqnarray}
P_{e\alpha} & = & A_{\alpha} + B_{\alpha}\cos\delta_{\rm CP} + C_{\alpha} \sin\delta_{\rm CP}, \nonumber  \\
P_{\alpha e} & = & A'_{\alpha} + B'_{\alpha} \cos\delta_{\rm CP} - C'_{\alpha} \sin\delta_{\rm CP}, \nonumber \\
P_{\bar{e}\bar{\alpha}} & = & \bar{A}_{\alpha} + \bar{B}_{\alpha}\cos\delta_{\rm CP} - \bar{C}_{\alpha} \sin\delta_{\rm CP} \label{eq:Biprob}   \\
P_{\bar{\alpha} \bar{e}} & = & \bar{A}'_{\alpha} + \bar{B}'_{\alpha} \cos\delta_{\rm CP} + \bar{C}'_{\alpha} \sin\delta_{\rm CP},  \nonumber
\end{eqnarray}
where $X_{\alpha},X'_{\alpha},\bar{X}_{\alpha},\bar{X}'_{\alpha}$ are coefficients depending on the mixing angles, mass-squared differences, matter potentials, and neutrino energy/baseline, for $X=A,B,C$, $\alpha=\mu,\tau$. This result is true for any matter potential as long as there are no new interactions. In vacuum, $X_{\alpha}=X'_{\alpha}=\bar{X}_{\alpha}=\bar{X}_{\alpha}'$, for $X=A,B,C$, $\alpha=\mu,\tau$. In symmetric matter, $X_{\alpha}=X'_{\alpha}$ and $\bar{X}_{\alpha}=\bar{X}_{\alpha}'$, for $X=A,B,C$, $\alpha=\mu,\tau$.

\begin{figure}[htbp]
\begin{center}
\includegraphics[width=\linewidth]{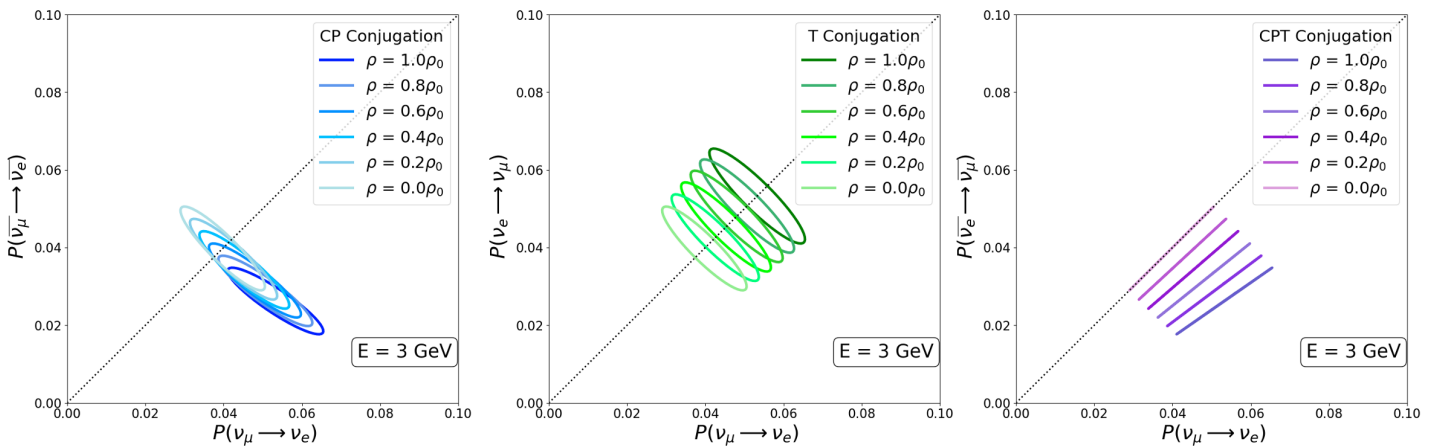}
\caption{Bi-probability comparisons for $L = 1300$ km and $E = 3$ GeV, characteristic for the upcoming DUNE experiment. Each panel depicts a different symmetry conjugation with $P(\nu_\mu \to \nu_e)$ common across all $x$-axes -- the left panel depicts CP conjugation, the center T conjugation, and the right CPT conjugation. In all cases, we assume the normal mass ordering and constant matter density along the path of propagation. Various levels of matter density are shown, between $0\%$ and $100\%$ of $\rho_0$, which is taken to be $2.84$ g/cm$^3$.\label{fig:Biprob_3GeV}}
\end{center}
\end{figure}

Bi-probability comparisons are performed to highlight the differences among related oscillation probabilities, including CP-conjugate channels ($P_{\alpha\beta} \leftrightarrow P_{\bar\alpha \bar\beta}$), T-conjugate channels ($P_{\alpha\beta} \leftrightarrow P_{\beta\alpha}$) and  CPT-conjugate channels ($P_{\alpha\beta} \leftrightarrow P_{\bar\beta \bar\alpha}$). As a first exploration, we consider the impact of nonzero matter density on the different lenses of a bi-probability comparison. This is demonstrated in~\cref{fig:Biprob_3GeV}, where we have adopted a DUNE-like situation as our point of reference -- we take $L = 1300$ km and $\rho_0 = 2.84$ g/cm$^3$. Each panel compares the oscillation probability at $E = 3$ GeV of $P(\nu_\mu \to \nu_e)$ to one conjugate channel -- CP (left), T (center), and CPT (right), where we keep all oscillation parameters fixed and vary $\delta_{\rm CP}$ to trace out the elliptical path. In each panel, with varying line darkness, we demonstrate the impact of modifying the (flat) matter density along the path of propagation, where the faintest (darkest) lines correspond to vacuum ($\rho = \rho_0$).

A great deal of interesting discussion exists regarding the CP-conjugate channel, e.g., the left panel of~\cref{fig:Biprob_3GeV}, so let us focus on the T- and CPT-conjugate ones. First, the CPT-conjugate regions in the right panel appear to all exhibit a line (or a very ``flat'' ellipse) in this parameter space. Given the expressions in  Eq.~(\ref{eq:Biprob}), one can calculate the area of the ellipse traced out by these paths when varying $\delta_{\rm CP}$. Using the analytic approximations for oscillation probabilities in Ref.~\cite{Freund:2001pn}, we can demonstrate that, for a constant-matter potential, the CPT-conjugate ellipse vanishes to order $s_{13}^2(\Delta m_{21}^2/\Delta m_{31}^2)^2$. We have verified that, for instance, by artificially inflating $\Delta m_{21}^2 \rightarrow 10 \times \Delta m_{21}^2$, the areas of the CPT-conjugate ellipses become visible.

In these figures, the distance a point exists away from the diagonal dashed line indicates the total amount of invariance violation present in the system. Especially when considering the CP-conjugate case, it is clear that (fixing all oscillation parameters except $\delta_{\rm CP}$), large CP-invariance violation effects can be conjured by changing $\delta_{\rm CP}$ or changing the amount of matter along the path of propagation. Put another way, for a fixed $\delta_{CP}$ (a specific point along the ellipse), significant CP-invariance violating effects are induced as $\rho = 0 \to \rho = \rho_0$. This is, incidently, the phenomenon by which DUNE and T2HK (as well as T2K and NOvA) aim to exploit in order to detemine the neutrino mass ordering. We do not comment on this here but present some results for both mass orderings in Appendix~\ref{app:FullBiprobability}.

Consistent with the discussion above, the interplay between these two effects is qualitatively different when considering T conjugation (the center panel of~\cref{fig:Biprob_3GeV}). A decent amount of TV is possible by varying $\delta_{\rm CP}$ in the vacuum case (faintest green ellipse), however when the matter density is increased, the span of these ellipses does not grow significantly. Additionally, if we fix $\delta_{\rm CP} = 0$ or $\pi$ (i.e. no intrinsic TV), these points stay fixed on the diagonal dashed line independent of $\rho$, where they do not in the CP-conjugate ellipses. These results are similar to statements made above: matter-induced TV is small, especially when considering the constant-potential case.

\begin{figure}[htbp]
\begin{center} 
\includegraphics[width=0.35\linewidth]{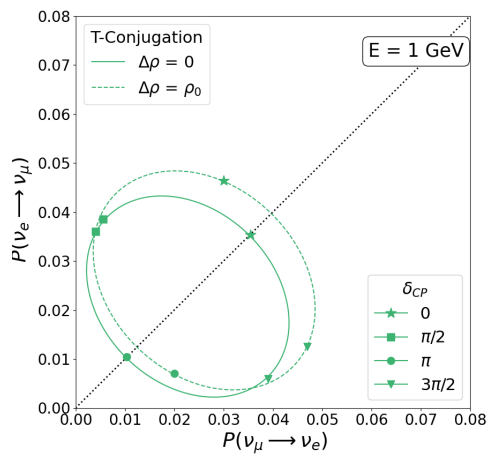}
\caption{T-conjugation comparison at $L = 1300$ km and $E = 1$ GeV for two matter density profiles: $\rho = \rho_0$ (2.84 g/cm$^3$) constant (solid) and $\rho = 0$ for $L/2$ and $2\rho_0$ for $L/2$ (dashed). In each ellipse, four values of the phase $\delta_{\rm CP}$ are labeled. See text for further detail. \label{fig:Biprob_Stepped}}
\end{center}
\end{figure}

Next, we consider the case of an asymmetric matter density profile, focusing on the T-conjugate bi-probability comparison. We demonstrate this for the DUNE baseline and $E = 1$ GeV in~\cref{fig:Biprob_Stepped}. Here, for illustrative purposes, we take a dramatic matter-density-profile comparison where either $\rho = \rho_0$ constant for the entire path of propagation (solid line), or one where $\rho = 0$ for the first $L/2$ and then $\rho = 2\rho_0$ for the remaining distance (dashed line). We also indicate explicitly four different points along the ellipses for $\delta_{\rm CP}$ corresponding to intrinsically CP-conserving and CP-violating parameters. We highlight here that the points in which CP (and T) are intrinsically conserved (star and circle) move from the diagonal to significantly away from it in the presence of a matter-density asymmetry. Additionally, the points of ``maximal'' CP-violation (square and triangle) are no longer the furthest points from the diagonal, e.g., there are other choices of $\delta_{\rm CP}$ that provide \textit{greater} T-invariance-violation in such a potential. In a nutshell, the asymmetric matter potential not only changes the size of the ellipse relative to what is observed with equivalent constant matter, it also ``rotates'' it.

\section{Concluding Remarks}
\label{sec:conclusion}
\setcounter{equation}{0}

Investigations of CP invariance and T invariance using neutrino oscillation experiments are impacted by matter effects. Matter effects on CP-odd oscillation observables (e.g., $P_{\mu e}-P_{\bar{\mu}\bar{e}}$) have been extensively explored in the literature. It is well known that matter-induced CP-invariance violation can be large for Earth-bound experiments and neutrino energies above hundreds of MeV. This is a robust claim; it does not depend too much on how the matter is distributed along the neutrino trajectory and does not significantly depend on whether CP invariance is intrinsically conserved. 

Here, we investigated the impact of matter effects on T-odd observables  (e.g., $P_{\mu e} - P_{e\mu}$). We made use of the quantum-mechanical formalism of neutrino-flavor evolution and attempted to be comprehensive and pedagogical. A summary of the main points we highlight is as follows. 
\begin{itemize}
\item Matter effects impact T-odd observables, but matter-induced T-invariance violation (TV) is qualitatively different from, and more subtle than,  matter-induced CP-invariance violation. 
\item If the matter distribution is symmetric relative to the neutrino production and detection points, matter effects will not introduce any ``extra'' TV. For example, if $P_{\mu e} = P_{e\mu}$ in vacuum, $P_{\mu e} = P_{e\mu}$ in symmetric matter. On the other hand, if there is intrinsic TV, matter effects can modify the size of the T-odd observable.  For example, if $P_{\mu e} \neq P_{e\mu}$ in vacuum, $P_{\mu e} \neq P_{e\mu}$ in matter, but $(P_{\mu e}-P_{e\mu})_{\rm vacuum}\neq (P_{\mu e}-P_{e\mu})_{\rm matter}$. For Earth-bound long-baseline oscillation experiments, we argued, mostly via concrete examples, that these effects are quantitatively small. 
\item If the matter distribution is not symmetric relative to the neutrino production and detection points, there is genuine matter-induced TV. For example, even if $P_{\mu e} = P_{e\mu}$ in vacuum, $P_{\mu e}$ and $P_{e\mu}$ may be different in the presence of asymmetric matter. For Earth-bound long-baseline oscillation experiments, we argued, mostly via concrete examples, that these effects are quantitatively very small. This remains true for unrealistically-asymmetric matter potentials (for example, we investigated the effects of ``hollowing out'' 50\% of the Fermilab-to-SURF neutrino trajectory). 
\end{itemize} 

We explored in detail the special case of two neutrino flavors. In this case, remarkably, all canonical T-odd observables (e.g., $(P_{\mu e}-P_{e\mu})$), for all neutrino energies and baselines, vanish independent of the matter potential or the presence of new interactions that violate T invariance. It would, of course, be possible to study whether T invariance was violated if there were only two neutrino flavors. The most straight-forward T-odd observable would be what we referred to as a proper test of T-invariance violation: comparing $P^{ab}_{\mu e}$ with $P^{ba}_{e\mu}$, where the superscripts refer to the neutrino production and detection points, respectively. If there is intrinsic TV, for two flavors, $P^{ab}_{\mu e}\neq P^{ba}_{e\mu}$ (as long as the matter potential is not symmetric!). 

Many of the results presented here are known but we believe our derivations and discussions help clarify and disseminate them accurately. In particular, we showed that $P_{ee}$ does not depend on $\delta$ or $\theta_{23}$ for any matter potential (constant, symmetric, asymmetric) as long as there are no interactions beyond those in the Standard Model. New interactions, including flavor-diagonal, CP-conserving interactions, invalidate this result. Similarly, we showed that, absent new interactions, bi-probability plots involving electron (or anti-electron) flavor define ellipses in the relevant bi-probability plane -- for any matter potential -- while bi-probability plots involving $P_{\mu\tau}$ and related channels do not (even in vaccum). 

T-odd observables are fascinating in their own right, but they are of limited practical use given currently available neutrino sources. With future intense, well-characterized, high-energy electron-neutrino beams, circumstances may chance dramatically. This would be the case if neutrino beams from high-energy muon storage rings -- so-called `neutrino factories' --  became available and allowed for precision studies of T-invariance violation in neutrino oscillations.

\section*{Acknowledgements}
OMB and AdG thank Woodkensia Charles and Grace Reesman for discussions and collaboration at the very early stages of this work. 
This work was supported in part by the US Department of Energy (DOE) grants \#de-sc0010143 and \#de-sc0010813 and in part by the National Science Foundation under Grant Nos.~PHY-1630782 and PHY-1748958. 

\appendix
\section{Biprobability Plots -- Mass-ordering Comparison}\label{app:FullBiprobability}
\begin{figure}[!htbp]
\begin{center}
\includegraphics[width=0.95\linewidth]{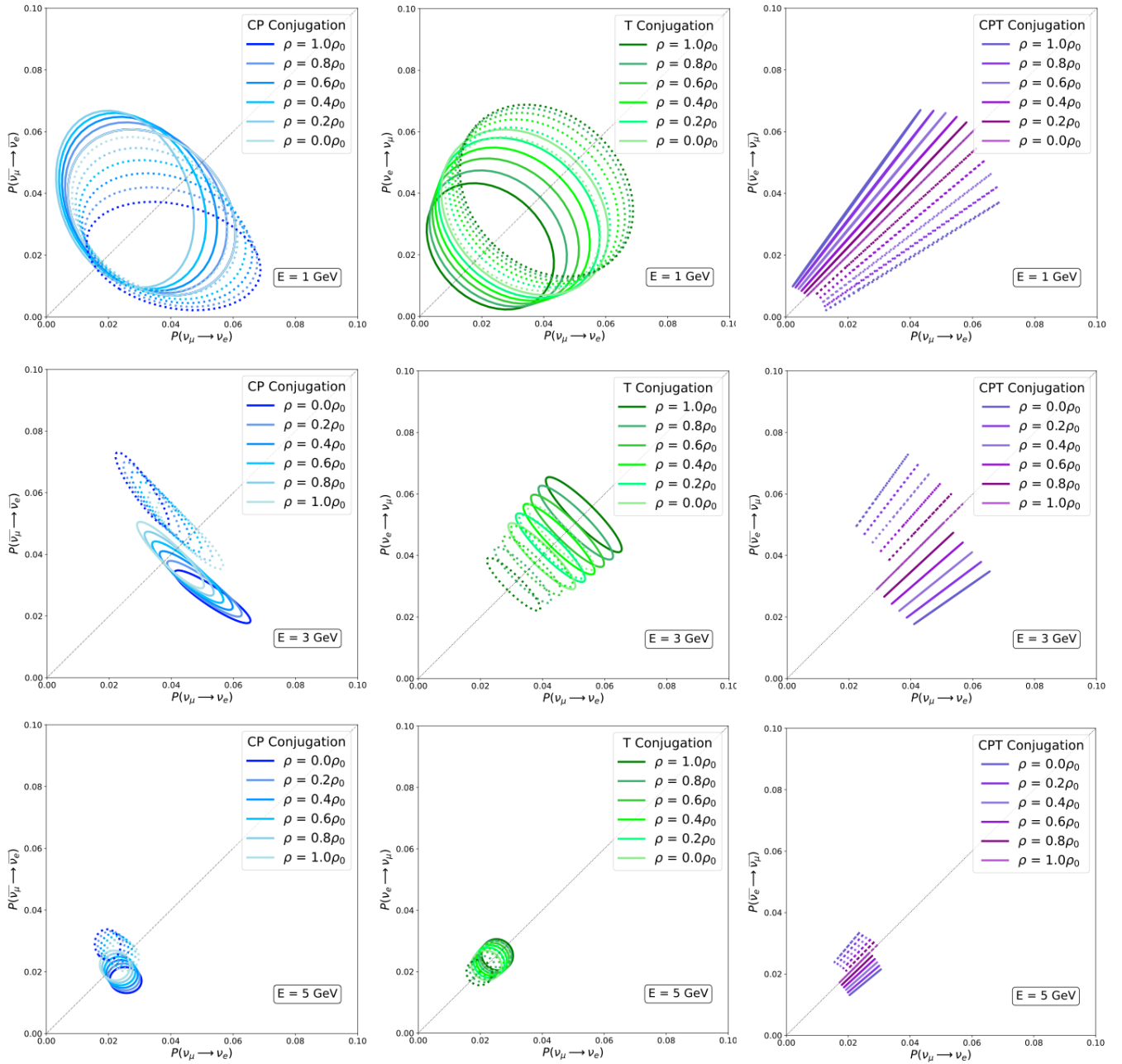}
\caption{Illustration of the impact of the neutrino mass ordering and energy in bi-probability comparisons of different conjugations: CP (left), T (center), and CPT (right). Each column corresponds to a different neutrino energy, where the baseline $L = 1300$ km is fixed in all. We vary the (constant) matter density along the path of propagation with different line darknesses, with $\rho = 0$ corresponding to the faintest lines and $\rho = \rho_0 = 2.84$ g/cm$^3$ corresponding to the darkest. In each case, we allow for the normal mass ordering (solid lines) and the inverted mass ordering (dashed), with oscillation parameters taken from Ref.~\cite{Esteban:2024eli}. \label{fig:BiprobNOIO}}
\end{center}
\end{figure}

In~\cref{subsec:Biprobability}, we discussed the utility of making comparisons of different conjugate channels, i.e. CP, T, and CPT, for long-baseline neutrino oscillation experiments. There (see~\cref{fig:Biprob_3GeV}) we focused on the DUNE experiment's baseline ($1300$ km) and characteristic energy ($3$ GeV), fixing the oscillation parameters such that the neutrino mass ordering is normal. In this appendix, we allow the neutrino energy and the mass ordering to vary. \cref{fig:BiprobNOIO} depicts a variety  of choices. In the top row, $E = 1$ GeV, in the middle row, $E = 3$ GeV, (in agreement with~\cref{fig:Biprob_3GeV}), and, in the bottom row, $E = 5$ GeV. The different columns, as in~\cref{fig:Biprob_3GeV}, correspond to different conjugations: CP (left), T (center) and CPT (right), from left to right. As in the main text, we are focused here on the impact of changing the (constant) matter-density-profile and the inferred symmetry violations. Different line darknesses indicate different matter densities, ranging from $\rho = 0$ to $\rho = \rho_0 = 2.84$ g/cm$^3$. Solid lines correspond to the normal mass ordering, while dashed lines correspond to the inverted mass ordering ($m_3 < m_1 < m_2$). We take the best-fit neutrino-oscillation parameters according to each mass-ordering separately according to the most recent results of Ref.~\cite{Esteban:2024eli}.

\bibliographystyle{utphys}
\bibliography{References}

\end{document}